\documentclass{PoS}

\usepackage{amssymb,fontenc,times,mathptmx,graphicx}

\usepackage{amsmath,latexsym}
\usepackage{mathrsfs}
\usepackage{bbm}
\title{Cutoff effects for Wilson twisted mass fermions
 at tree-level of perturbation theory}

\ShortTitle{Cutoff effects for Wilson twisted mass fermions
 at tree-level of perturbation theory}

\author{
\vspace*{-10mm}

\begin{flushright}
HU-EP-07/55\\
SFB/CPP-07-66\\
DESY 07-181\\
\end{flushright}
\vspace*{7mm}
K. Cichy$^P$, \speaker{J. Gonz\'alez L\'opez}$^{BZ}$,
K. Jansen$^Z$, A. Kujawa$^P$ and A. Shindler$^N$\\
\llap{$^B$}Humboldt--Universit\"at zu Berlin, Institut f\"ur Physik,
Newtonstrasse 15, 12489 Berlin, Germany\\
\llap{$^P$}Adam Mickiewicz University of Poznan, Faculty of Physics,
Umultowska 85, 61-614 Poznan, Poland\\
\llap{$^Z$}DESY, Platanenallee 6, 15738 Zeuthen, Germany\\
\llap{$^N$}NIC, Platanenallee 6, 15738 Zeuthen, Germany\\
E-mail: \email{kcichy@epf.pl}, \email{jenifer.gonzalez.lopez@desy.de}
\email{Karl.Jansen@desy.de},
\email{agnieszkakujawa@gazeta.pl},
\email{andrea.shindler@desy.de}}

\abstract{We study cutoff effects
at tree-level of perturbation theory
for standard Wilson and Wilson twisted mass fermionic lattice actions
with $N_{f}=2$ flavour degenerate quarks.
The discretization effects are investigated by computing the mass spectrum and
decay amplitudes for different hadron interpolating fields and the
scaling behaviour towards the continuum limit is analyzed.
It is shown that the Wilson and the mass average methods are equivalent and
lead 
to $O(a)$ improved $\mathcal{R}_{5}$-parity even lattice observables.
We also demonstrate that automatic $O(a)$ improvement works in case of Wilson
twisted mass fermions at maximal twist and that this improvement is realized even if
the condition of maximal twist is achieved only up to $O(a)$ cutoff effects.
We demonstrate that in the chiral limit standard Wilson fermions show 
scaling violations of $O(a^2)$ while for 
maximally twisted mass fermions these violations 
are only of $O(a^4)$. 
For our analytical calculations, lattices with sizes $L=aN$ and
periodic boundary conditions
in the spatial directions have been chosen while infinite extension in
the time direction, $L_{4}=\infty$, is considered.}

\FullConference{The XXV International Symposium on Lattice Field Theory\\
		 July 30-4 August 2007\\
		 Regensburg, Germany}

\begin{document}
\section{Introduction: Wilson twisted mass action}

\noindent In this contribution, 
we study the cutoff effects
of observables computed on a lattice with lattice spacing $a$  
when Wilson twisted mass fermions at
maximal twist are considered \footnote{The interested reader may find 
a much detailed account of this work in the thesis of J.~G.~L., see
ref.~\cite{Jenithesis}}. 
The twisted mass QCD action in the continuum,
at tree-level of perturbation theory, is given by
\begin{equation}\label{eq:and21}
S_{F}^{tm}[\chi,\bar{\chi}]=\int\, d^{4}x\, \bar{\chi}(x)\, 
(\gamma_{\mu}\partial_{\mu}+m_{0}+i\mu_{q}\gamma_{5}\tau^{3})\, \chi(x),
\end{equation} 
with $\tau^{3}$ the third Pauli matrix acting in flavour space
and $\{\chi,\bar{\chi}\}$ the so-called twisted basis.
In equation~\eqref{eq:and21}, 
$m_{0}$ denotes the bare \emph{untwisted quark mass} and
$\mu_{q}$ the bare \emph{twisted quark mass}. The mass term
can be written in terms of a
polar mass $M$ and a polar angle $\alpha$ as
\begin{equation}\label{eq:and24}
m_{0}+i\mu_{q}\gamma_{5}\tau^{3}=M\, e^{i\alpha\gamma_{5}\tau^{3}}\qquad \text{with} \qquad
M=\sqrt{m_{0}^{2}+\mu_{q}^{2}}\, ,\quad \alpha=\arctan (\frac{\mu_{q}}{m_{0}}).
\end{equation}
The twisted basis $\{\chi,\bar{\chi}\}$
is related to the so-called physical basis $\{\psi,\bar{\psi}\}$
by the non-anomalous axial transformation
\begin{equation}\label{eq:and26}
\psi(x)=e^{i\omega\gamma_{5}\frac{\tau^{3}}{2}}\chi(x) \qquad
\bar{\psi}(x)=\bar{\chi}(x)e^{i\omega\gamma_{5}\frac{\tau^{3}}{2}}\; ,
\end{equation}
for the particular choice of the twisting angle $\omega=\alpha$,
since it brings the twisted mass QCD action back to the standard form.
\noindent The Wilson-regularized twisted mass action (Wtm)
written in the twisted basis has the form
\begin{equation}\label{eq:SQCDWFtm}
S_{F}^{Wtm}[\chi,\bar{\chi}]=a^{4}\sum_{x}\, \bar{\chi}(x)\, 
\lbrack D_{W}+m_{0}+i\mu_{q}\gamma_{5}\tau^{3}\rbrack \, \chi(x)\, ; \quad
D_{W}=\frac{1}{2}\lbrace \gamma_{\mu}(\partial_{\mu} 
+\partial_{\mu}^{*})-ar\partial_{\mu}^{*}\partial_{\mu} \rbrace
\end{equation}
with $D_{W}$ the Wilson-Dirac operator of the free case,
$\partial_{\mu}$ and $\partial_{\mu}^{*}$ the forward and backward 
partial lattice derivatives and
$-\frac{ar}{2}\partial_{\mu}^{*}\partial_{\mu}$ the Wilson term.

\section{Wilson twisted mass free-fermion propagator with infinite time-extent lattices}

\noindent The expression for the Wilson twisted mass fermion propagator
in the twisted basis, at tree-level of perturbation theory (PT) and in momentum space is
given by
\begin{equation}\label{tmprop}
\widetilde{S}(p)= \frac{-i \frac{1}{a}\sin (ap_{4})\gamma ^{4}\mathbbm{1}_{f}-i \mathcal{K}\mathbbm{1}_{f} +  \bigl[ \frac{r}{a}\left( 1 - \cos (ap_{4}) \right) + \mathcal{M} \bigr]\mathbbm{1}\mathbbm{1}_{f}-i\mu _{q}\, \gamma ^{5}\tau ^{3} }{ \frac{1}{a^{2}}\sin ^{2} (ap_{4}) + \mathcal{K}^{2} + \bigl[ \frac{r}{a} \left( 1 - \cos (ap_{4}) \right) + \mathcal{M} \bigr] ^{2}+\mu _{q}^{2}},
\end{equation}
where 
$\mathbbm{1}$ and $\mathbbm{1}_{f}$ are the identity matrices in Dirac
and flavour space. The structure in colour space has not been written
since it is just an identity matrix at tree level of PT and we have defined, 
\begin{equation}\label{eq:wilmass}
\mathcal{M}(p)=m_{0}+\frac{2r}{a}\sum_{i=1}^{3}\, \sin^{2}(\frac{ap_{i}}{2}), \qquad
\mathcal{K} = \frac{1}{a}\sum_{i=1}^{3}\, \gamma^{i} \sin(ap_{i}).
\end{equation} 

\noindent We obtain the expression for the quark propagator in the time-momentum
representation when an infinite extension of the time direction is considered.
To perform the integral in the continuous
momentum $p_4$ 
amounts to performing a contour integral and compute the residues
of the integrand. The integration contour
encloses 
the poles of the integrand at the energy points
\begin{displaymath}
\left\{ \begin{array}{lll}
p_{4} =\frac{1}{a}iE_{1}\qquad
&\cosh E_{1}= \frac{\mathcal{U} - r \left( r + a\mathcal{M} \right)}{1 - r^{2}}\quad
&\text{physical pole}\\
p_{4}= \frac{1}{a}(\pi + iE_{2})\qquad&
\cosh E_{2}= \frac{\mathcal{U} + r \left( r + a\mathcal{M} \right)}{1 - r^{2}}\quad
&\text{doubler pole}
\end{array}\right.
\end{displaymath}
with $\mathcal{U} ^{2}= \left( 1 + ar\mathcal{M} \right) ^{2} +
\left( 1 - r^{2} \right) \left( a^{2}\mathcal{K} ^{2} + a^{2}\mathcal{M} ^{2} + a^{2}\mu _{q}^{2}\right)$.
The final expression for the propagator,
in the limit $r\rightarrow 1$, is then\footnote{This expression
is valid for all possible values of
the discrete Euclidean time $t$, including if it is negative or zero.
The function $\text{sgn}(t)$
is the sign of t,
and we have denoted $\text{sgn}(0)\equiv 0$.
It is just a convention in order to give one general
expression for the propagator for all possible values of $t$.},
\begin{equation}\label{eq:propinftm1}
\begin{split}
S_{\infty} \bigl( \vec{p},t \bigr)= \frac{1}{2\,\mathcal{U} \sinh E_{1}}&\Bigl
\{\text{sgn}(t)\sinh E_{1} \gamma _{4}\mathbbm{1}_{f} -ia\mathcal{K}\mathbbm{1}_{f}
+ \bigl[ \left( 1 - \cosh E_{1} \right) + a\mathcal{M}\bigr]\mathbbm{1}\mathbbm{1}_{f}\\
&- ia\mu _{q}\gamma _{5}\, \tau ^{3}\Bigr\}e^{- E_{1}|\frac{t}{a}|}\:
+\: \delta _{\frac{t}{a},0}\, \frac{1}{2\,(1 + a\mathcal{M})}\mathbbm{1}\mathbbm{1}_{f}
\end{split}
\end{equation}
where $\cosh E_{1} = 1+ \frac{a^{2}\mathcal{K}^{2} + a^{2}\mathcal{M}^{2} + a^{2}\mu _{q}^{2}}{2\,(1 + a\mathcal{M})}$.\\

\noindent The fermion propagator is a matrix 
in Dirac space and can hence be 
decomposed in terms of the Dirac gamma matrices as  $S_{\infty} \bigl( \vec{p},t \bigr) = S_{U} \bigl(\vec{p},t \bigr)\,\mathbbm{1} +
\sum_{\mu}\,S_{\mu} \bigl(\vec{p},t \bigr)\,\gamma_{\mu} + S_{5} \bigl(\vec{p},t \bigr)\,\gamma _{5}$.
\section{Hadron correlation functions}

\subsection{Pseudo-scalar meson}
\label{sec:psFTM}

\noindent The interpolating fields describing the charged pions, 
$\pi^{+}$ and $\pi^{-}$
respectively, in the physical and the
twisted bases are
\begin{equation}
\mathcal{P}^{\pm}(x)\equiv \mathcal{P}^{1}(x)\mp i\mathcal{P}^{2}(x)=
P^{1}(x)\mp iP^{2}(x)
\end{equation}
where $\mathcal{P}^{a}(x)=\bar{\psi}(x)\gamma_{5}\frac{\tau^{a}}{2}\psi(x)$, with 
$a=1,2,3$, is the
pseudo-scalar density written in the physical basis
while $P^{a}(x)=\bar{\chi}(x)\gamma_{5}\frac{\tau^{a}}{2}\chi(x)$ is the 
pseudo-scalar density
written in the twisted basis.

\noindent The time dependence of the two-point correlation function
for the charged pseudo-scalar meson in the time-momentum representation
is given by
\begin{equation}\label{charpionCF}
C_{\mathcal{P}^{\pm}\mathcal{P}^{\pm}} (t)=
\frac{N_{c}\,N_{d}}{L^{3}}\sum_{\vec{p}}
\Bigl \{ |S^{u}_{U}(\vec{p},t)|^{2} + \sum_{\mu =1}^{4}\:
|S^{u}_{\mu}(\vec{p},t)|^{2} +|S^{u}_{5}(\vec{p},t)|^{2} \Bigr \}. 
\end{equation}
We denote the Wilson twisted mass fermion propagator in the
twisted basis
for ``$u$ quarks'' as $S^{u}$. Note that this is not identical 
in the
twisted mass case with the propagator for ``$d$ quarks''
denoted as $S^{d}$. $N_{c}$ ($N_{d}$) is the number of colours (Dirac components).

\subsection{Proton}
\label{sec:proFTM} 

\noindent The local interpolating field describing the proton in both bases is
given by
\footnote{The Greek (Latin) letters denote Dirac
(colour) components and $u$, $d$ denote the flavour content. The notation used for the 
flavour structure is
$\psi=\binom{\hat{u}}{\hat{d}}$ and $\chi=\binom{u}{d}$. 
$C$ is the charge conjugation matrix
and $[\:]$ denotes spin trace.},
\begin{equation}\label{eq:defproton}  
\mathscr{P}_{\alpha}(x)\equiv -\sqrt{2}\epsilon _{abc}\bigl[\bar{\hat{d}}_{a}^{T}(x)\,C^{-1}
\gamma _{5} \hat{u}_{b}(x)\bigr]\, \hat{u}_{\alpha ,c}(x)=
\sqrt{2}\,\epsilon _{abc}\,\bigl[ d_{a}^{T}(x)\,C^{-1}\gamma _{5}\, u_{b}(x)\bigr]\,
e^{i\frac{\omega}{2} \gamma_{5}} \, u_{\alpha ,c}(x). 
\end{equation}
\noindent The expression for the time dependence of the proton correlation function is
then
\begin{equation}\label{eq:projectionP2}
C_{\mathscr{P}\bar{\mathscr{P}}}(t)=\frac{N_{c}N_{d}}{L^{6}}\,\sum_{\vec{p}}\sum_{\vec{q}}\Bigl \{
\cos(\omega)\, L_{U}(\vec{p},\vec{q},t)+L_{4}(\vec{p},\vec{q},t)+
i\sin(\omega)L_{5}(\vec{p},\vec{q},t)\Bigr \},
\end{equation}
with the definitions
\begin{equation}
\begin{split}
L_{U,5}(\vec{p},\vec{q},t)\equiv S^{u}_{U,5}(-(\vec{p}+\vec{q}),t)
\Bigl\{&(N_{d}+1)\,S^{u}_{U}(\vec{p},t)S^{u}_{U}(\vec{q},t)+(N_{d}+3)\,
\sum_{\mu =1}^{4}\,S^{u}_{\mu}(\vec{p},t)S^{u}_{\mu}(\vec{q},t)\\
&-(N_{d}+1)\,S^{u}_{5}(\vec{p},t)S^{u}_{5}(\vec{q},t)\Bigr\}
\end{split}
\end{equation}
\begin{equation}
\begin{split}
L_{\mu}(\vec{p},\vec{q},t)\equiv S^{u}_{\mu}(-(\vec{p}+\vec{q}),t)
\Bigl\{&(N_{d}+3)\,S^{u}_{U}(\vec{p},t)S^{u}_{U}(\vec{q},t)+(N_{d}+1)\,
\sum_{\mu =1}^{4}\,S^{u}_{\mu}(\vec{p},t)S^{u}_{\mu}(\vec{q},t)\\
&-(N_{d}+3)\,S^{u}_{5}(\vec{p},t)S^{u}_{5}(\vec{q},t)\Bigr\}.
\end{split}
\end{equation}

\section{Scaling test}

\subsection{Wilson average (WA) and mass average (MA) for standard Wilson fermions }
\label{sec:wa}

\begin{figure}
\vspace{-1cm}
\hspace{-1.5cm}
\includegraphics[width=0.4\textwidth,angle=270]{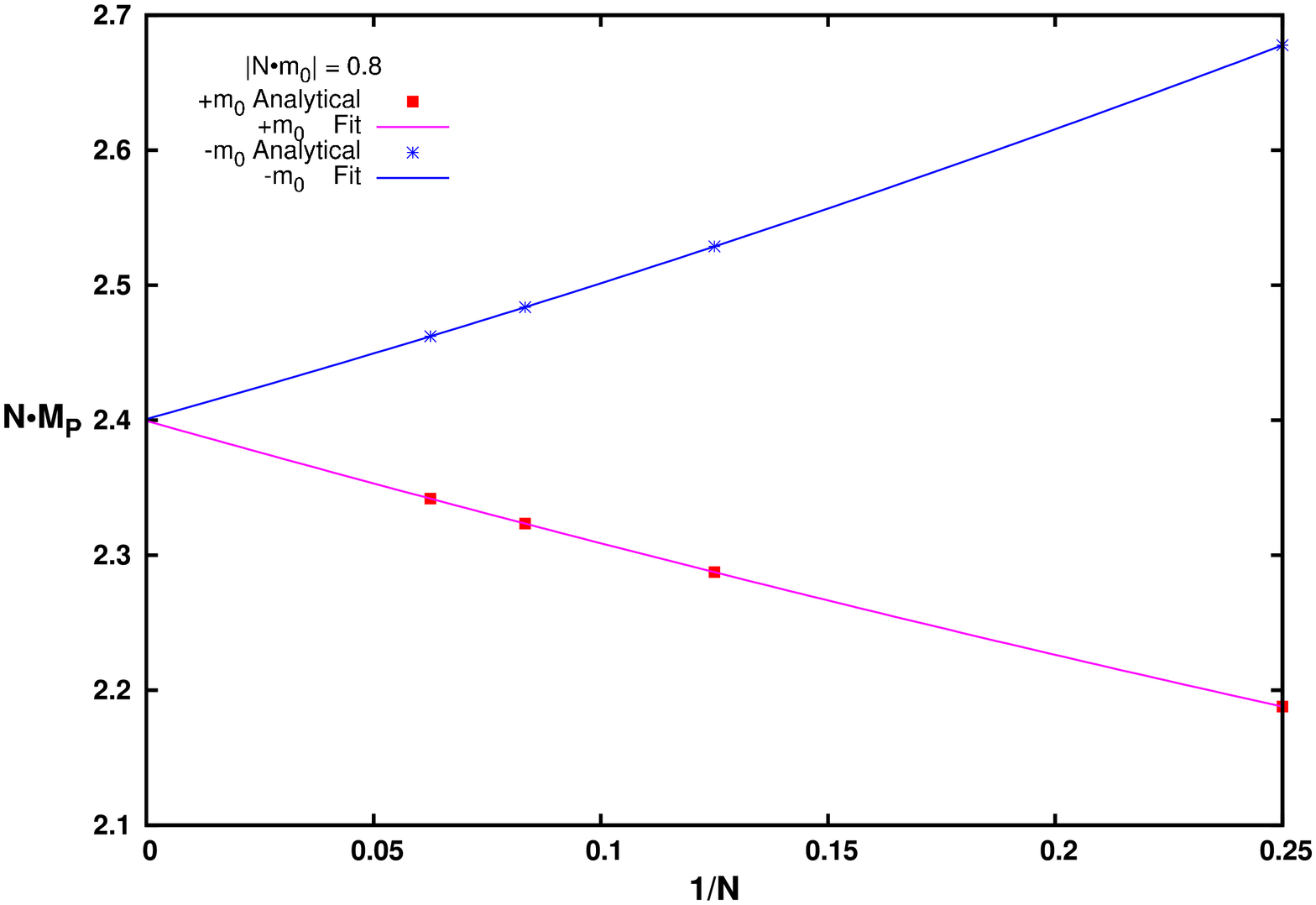}
\vspace{-1cm}
    \includegraphics[width=0.4\textwidth,angle=270]{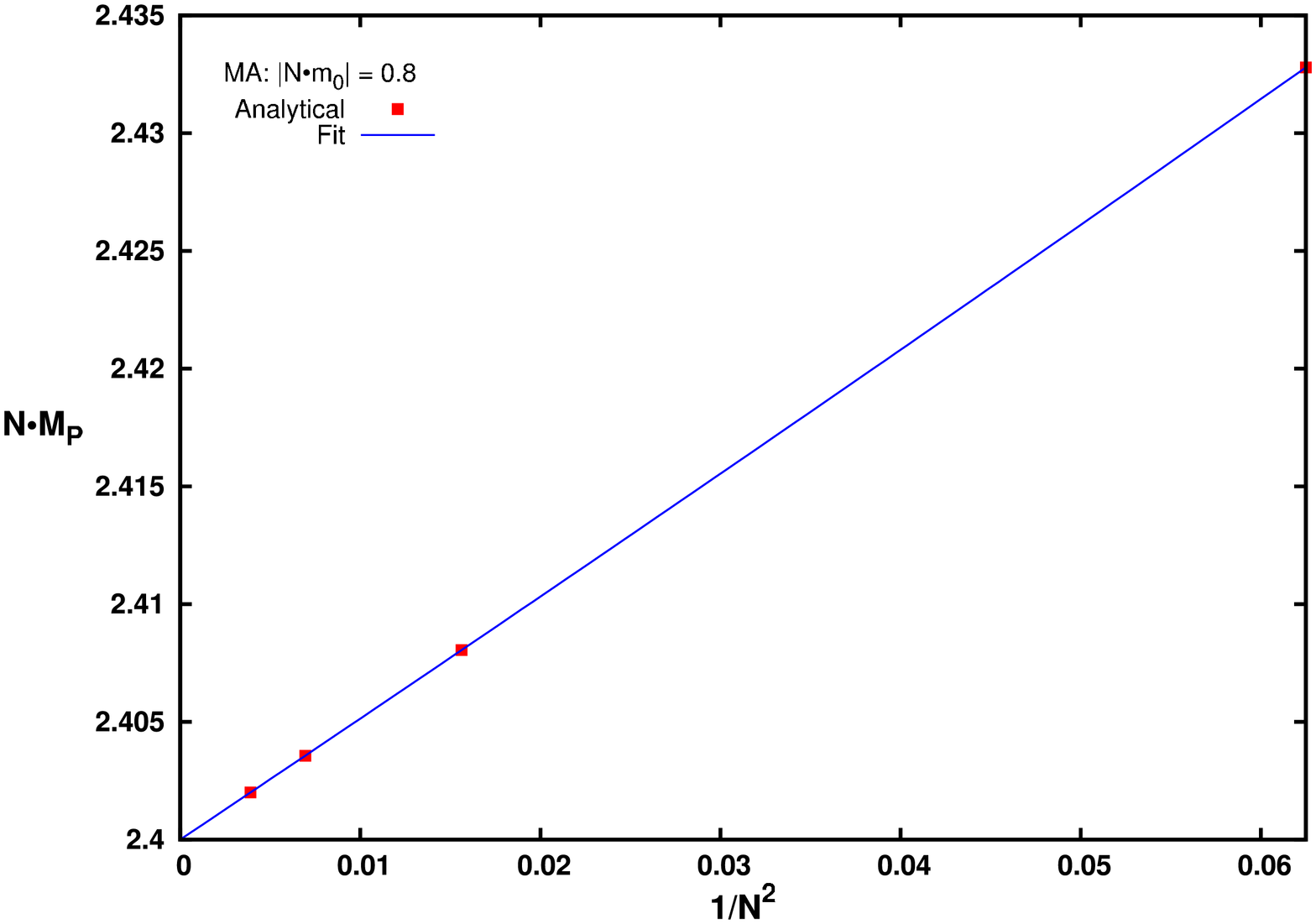}
\vspace{1cm}
\caption{In the left graph, the cutoff effects and 
the continuum limit of the proton mass
obtained from two standard Wilson actions differing only
in the sign of the quark mass, $|Nm_{0}|=0.8$ are shown.
The lattices are $4\leq N\leq 20$.
In the right graph, the average of the proton masses obtained from the same
two standard Wilson regularizations with quark masses $Nm_0=\pm 0.8$ (MA)
has been calculated.\label{fig:mapm1}}
\end{figure}

\noindent In Ref.~\cite{frezz} it has been demonstrated that 
when averaging physical observables computed with Wilson actions
having opposite signs of the
quark mass (MA) or opposite signs of the 
Wilson parameter (WA), these quantities are $O(a)$ improved.
Since WA and MA are equivalent, we show in Figure~\ref{fig:mapm1}, only
the cutoff effects at the example of the proton mass when the MA is performed.
In the left graph the behaviour of the proton mass $NM_{P}$
as a function of $\frac{1}{N}$ is given when 
a standard Wilson regularization is used.
In order to describe the behaviour of the physical quantities 
computed analytically at selected values of $\frac{1}{N}$, we use the following 
fitting functions:
\begin{equation}\label{eq:wfit}
y_1=a_{0}+a_{1}\frac{1}{N}+a_{2}\frac{1}{N^{2}}\qquad
y_2=b_{0}+b_{1}\frac{1}{N^{2}}+b_{2}\frac{1}{N^{4}}.
\end{equation}
Here $y_1$ ($y_2$) is the physical observable under consideration
and its value in the continuum limit is given by the coefficient
$a_0$ ($b_0$).
We use two functional forms, the first formula of
equation~(\ref{eq:wfit}) for a leading $\frac{1}{N}$ behaviour (standard Wilson fermions) 
and the second formula for $O(a)$-improved quantities.\\

\noindent The two lines in the left graph of Figure~\ref{fig:mapm1}
originate from a fit to equation~(\ref{eq:wfit}) and 
correspond to the proton mass
obtained from the same Wilson actions differing only in
the sign of the quark mass. 
The linear behaviour in $\frac{1}{N}$ shows the
$O(a)$ scaling violations present in the standard Wilson theory.
From the plot it is clear that in both cases
the value of the proton mass in the continuum limit is the same and
the expected one at tree-level of PT.
From the fit, the corresponding coefficients $a_1$ turn out to be 
the same in magnitude but have 
opposite signs for positive and negative quark masses.
Thus, performing the (MA), it is to be expected 
that the $O(a)$ effects cancel and the scaling behaviour changes 
drastically from a $\frac{1}{N}$ to 
a $\frac{1}{N^2}$ behaviour. This can indeed be seen in the right 
graph of Figure~\ref{fig:mapm1}.
Inspecting the fit coefficients $a_2$ and $b_1$,  
we find $a_{2}^{sW}\approx b_{1}^{MA}\approx 0.5$. 
Therefore, 
the magnitude of the leading order cutoff effects
does not only change from an $O(a)$ to an $O(a^{2})$ behaviour but also
the $O(a^{2})$ do not increase on performing the Wilson average with respect to the
standard case. 

\subsection{Wilson twisted mass fermions at maximal twist }
\label{sec:wtm}

\begin{figure}
\begin{center}
\vspace{-1cm}
    \includegraphics[width=0.4\textwidth,angle=270]{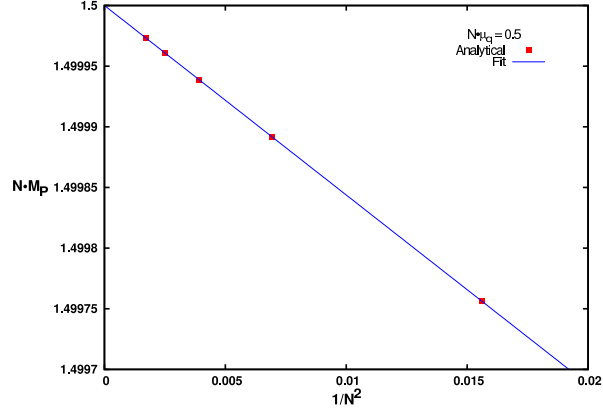}
\caption{Cutoff effects in the proton mass with Wtm fermions at maximal twist.
To study the case of maximal twist, we set the 
quark mass to $N\mu_{q}=0.5$. The lattices considered are taken as 
$4\leq N \leq 24$.\label{fig:tmpm}}
\end{center}
\end{figure}

\noindent Instead of performing a MA or WA, a way to obtain 
an {\em automatic} $O(a)$ improvement is to work with Wtm fermions
at maximal twist~\cite{frezz}.
Maximal twist is reached for a value of the twist angle of $\omega=\pi/2$. At tree-level 
of perturbation theory this 
can be achieved by simply setting the untwisted quark mass $m_0=0$.
The value of the quark mass is now fully given by the twisted quark mass,
$NM=N\mu_{q}$.
The proton mass as a function of $\frac{1}{N^{2}}$
is shown in Figure~\ref{fig:tmpm} and the corresponding fit is performed using 
the second fitting function of equation~(\ref{eq:wfit}).
The figure clearly demonstrates that \emph{automatic} $O(a)$ improvement is 
indeed working and that 
the cutoff effects have changed from a $\frac{1}{N}$ behaviour of
standard Wilson fermions to a $\frac{1}{N^{2}}$
behaviour when maximally twisted mass Wilson fermions are employed.
Moreover, as a result of the fit, the coefficient $b_1$ comes out to be 
very small,  $b_{1}^{tm}=O(10^{-2})$. This 
value is 
one order of magnitude smaller than the corresponding coefficient 
$a_2$ of the $O(a^2)$ effects for standard Wilson fermions which we find to be 
$a_{2}^{sW}=O(10^{-1})$. Note the the value of $b_1$ is also smaller than 
the one for the case 
of MA discussed above.\\

\noindent We have also analyzed the chiral limit behaviour 
of the pion and proton masses by determining the coefficients $a_1$ and $b_1$ 
of equations~(\ref{eq:wfit}).
In the case of standard Wilson fermions, the coefficient $a_1$
which determines the size 
of the $O(a)$ cutoff effects vanishes in the chiral limit
thus leading to only $O(a^2)$ scaling violations in the massless theory.
For Wtm fermions at maximal twist the situation is even better.
Here, the coefficient  
of the $O(a^{2})$ cutoff effects vanishes in the chiral limit,
thus leading to scaling violations of $O(a^{4})$ only  
since all odd powers of $a$ vanish for maximally twisted mass fermions.
Therefore, for Wilson twisted mass fermions at maximal twist 
the breaking of chiral symmetry
at finite lattice spacing is 
much smaller than for standard Wilson fermions and maximally twisted mass
fermions are indeed chirally improved.
\subsubsection{Out of maximal twist}
\label{sec:omt}

\noindent Here we want to study a situation 
when we allow an $O(a)$ error in 
setting the untwisted quark mass to zero. In order to realize this 
situation at tree-level of PT
we `force' these effects by simply
fixing the twisted mass to be the physical quark mass
and the untwisted mass is set to be proportional to $\frac{1}{N}$, as
$N\mu_{q}=\alpha$ and $Nm_{0}=\frac{\beta}{N}\backsim O(a)$
where $\alpha$ is kept fixed
and $\beta$ is a measure parametrizing the amount of violation
of the maximal twist setup. The 
twist angle $\omega$ and the bare polar mass $M$ can be
obtained as a function of $\alpha$ and $\beta$ as
\begin{equation}\label{eq:womt}
\omega = \frac{\pi}{2}-\Bigl (\frac{\beta}{\alpha}\Bigr )\frac{1}{N}+ O(\frac{1}{N^2}), \qquad
NM= \alpha \Bigl
[ 1+\frac{1}{2}\Bigl (\frac{\beta}{\alpha}\Bigr )^{2}\frac{1}{N^{2}}+O(\frac{1}{N^4})\Bigr ].
\end{equation}
\noindent Therefore, even if the condition of maximal twist
can be only obtained up to $O(a)$ cutoff effects, 
which is generically the case in practical numerical simulations, 
the observables, which are only functions of the polar mass, are still
automatically $O(a)$ improved.\\

\begin{figure}
\vspace{-1cm}
\hspace{-1.5cm}
    \includegraphics[width=0.4\textwidth,angle=270]{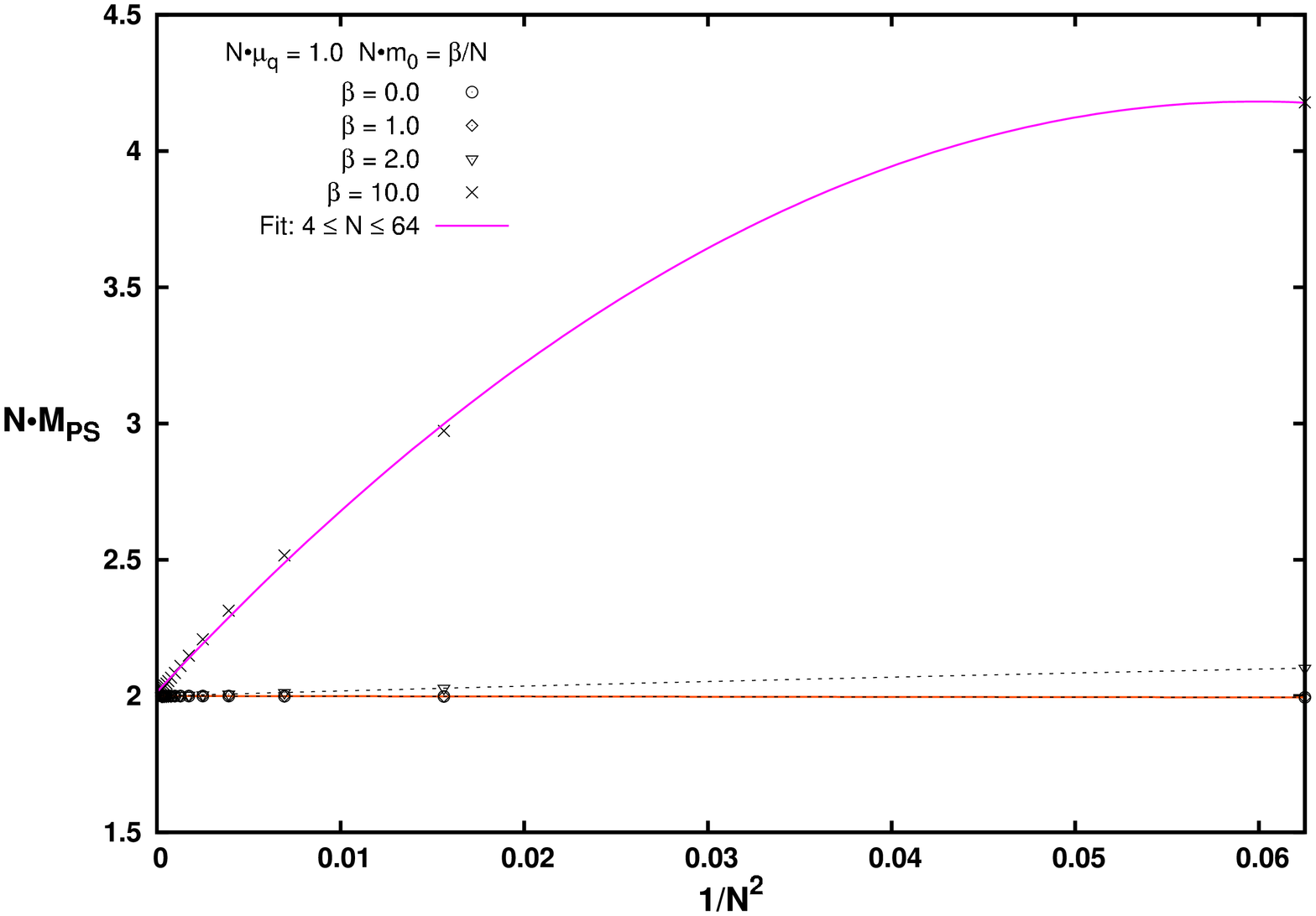}
\hspace{0.25cm}
\vspace{-1cm}
    \includegraphics[width=0.4\textwidth,angle=270]{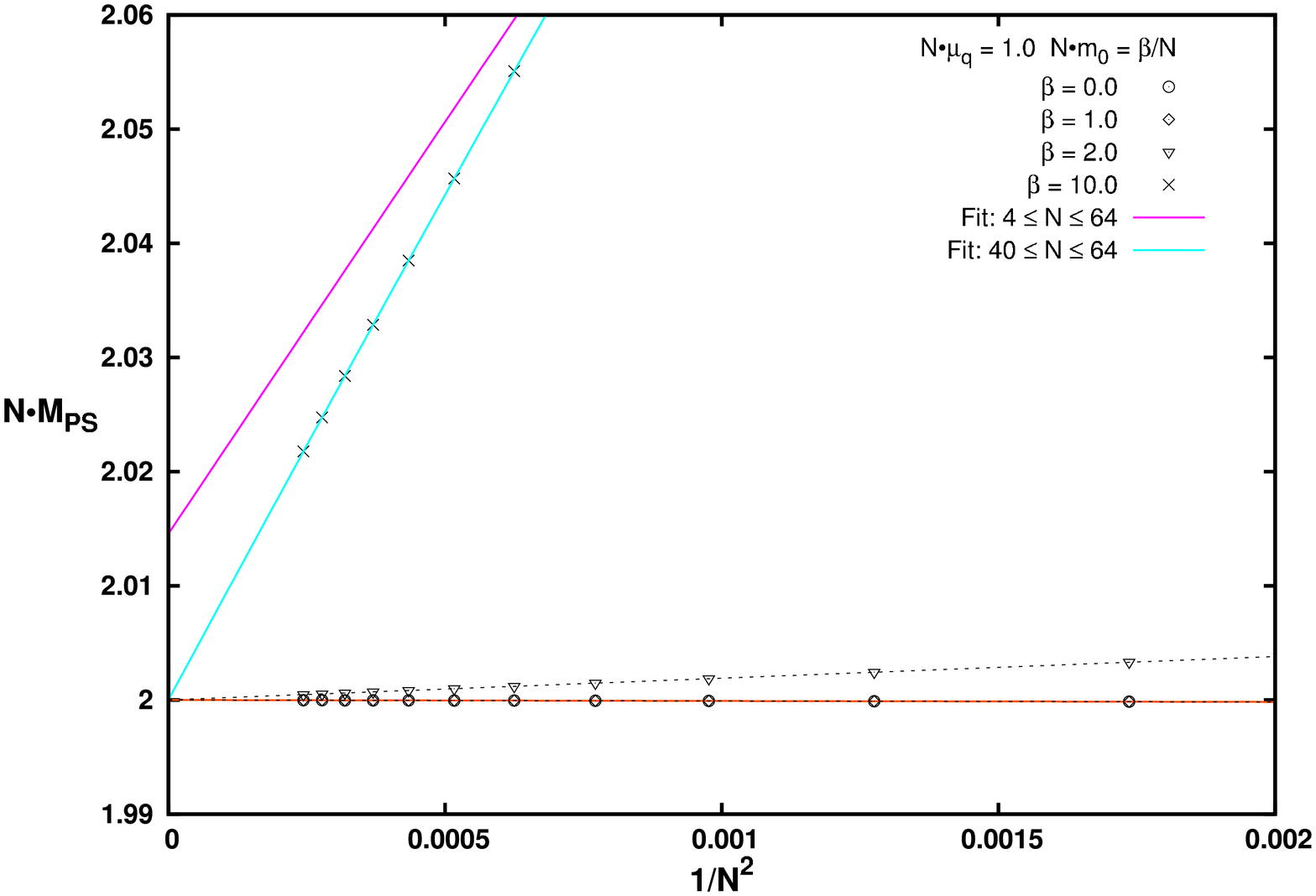}
\vspace{1.0cm}
\caption{Left graph: Behaviour of the pion mass as a function of $\frac{1}{N^{2}}$,
for lattices with size $4\leq N \leq 64$.
The twisted quark mass is set to $N\mu_{q}=1.0$ and the untwisted quark mass
is zero up to $O(a)$ cutoff effects i.e.
$Nm_{0}=\frac{\beta}{N}$ with $\beta = 0.0,1.0,2.0,10.0$.
Right graph: a zoom of the graph on the left with an additional fit for
the analytical data corresponding to $\beta =10.0$ which considers only large
lattices $40\leq N \leq 64$.\label{fig:outmt2}}
\end{figure}

\noindent Moreover,
equation~\eqref{eq:womt} also shows how the
size of the leading discretization effects
depends on the ratio between the untwisted
and twisted quark masses. This ratio in turn determines 
the value of the lattice spacing at which the asymptotic $\frac{1}{N^2}$ scaling 
sets in. Only when this ratio is small enough and hence the lattice 
does not need to be chosen too large a 
reliable continuum limit using reasonably sized lattices can be performed.
The left graph of Figure~\ref{fig:outmt2} demonstrates that 
the asymptotic scaling sets in for lattices with $4\leq N \leq 64$
when $0\lesssim \frac{\beta}{\alpha}\lesssim 2$.\\

\noindent However, for $\frac{\beta}{\alpha}\,\gtrsim 10$
the continuum limit is not
reliable anymore if $N$ is chosen to be too small.
This can be seen in the right graph of Figure~\ref{fig:outmt2}.
Using only small values of $N$ leads 
to an inconsistent continuum limit value.
Therefore, larger lattices are needed in order to obtain
the correct continuum behaviour as
can be also seen in the right graph of Figure~\ref{fig:outmt2}.
Here we have added a fit
of the data for a value of $\beta= 10.0$ taking only large lattices 
into account, 
i.e. using only values of $N \geq 40$.
In this case, indeed  
the right continuum value is obtained.
Of course, for practical simulations, using only lattices with $N \gg 40 $ appears to be 
rather unrealistic.

\vspace{-0.5cm}
\section{Conclusions}
\label{sec:conclusions}

\noindent In this contribution we have demonstrated at tree-level of perturbation theory
that when Wtm fermions at maximal twist are considered,
physically relevant quantities are automatically $O(a)$ improved.
In addition, the magnitude of the leading $O(a^2)$ corrections is rather small.
In the chiral limit the $O(a^{2})$ ($O(a)$) effects disappear leading
then to scaling violations of $O(a^{4})$ ($O(a^2)$) in the case of Wtm 
fermions at maximal twist (Wilson fermions).
Therefore, Wtm fermions at maximal twist show a substantially improved scaling 
and chiral behaviour when compared to standard Wilson
fermions which render Wtm fermions a powerful
formulation of lattice QCD.

\vspace{-0.5cm}
%Acknowledgments
\acknowledgments

\noindent 
We  want to thank
M. Brinet, V. Drach and C. Urbach
for help cross-checking some of the results presented here. 
We are also grateful to them and to M. M\"uller-Preussker for valuable 
discussions and comments.
J.~G.~L. thanks the SFB-TR9 for the financial support.

%Bibliography
%\bibliography{latt_07}

\end{document}